# Deep Camera: A Fully Convolutional Neural Network for Image Signal Processing

**Sivalogeswaran Ratnasingam**
ON Semiconductor, 5005 East McDowell Road, Phoenix, AZ 85008 USA

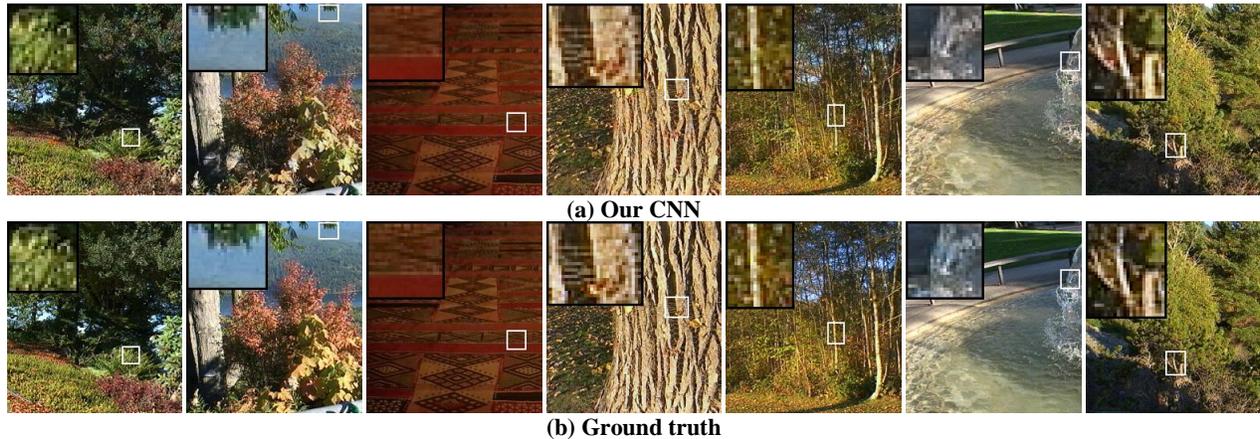

**(a) Our CNN**

**(b) Ground truth**

**Figure 1:** Our convolutional neural network performs the entire processing involved in an image signal processing (ISP) pipeline including denoising, white balancing, exposure correction, demosaicing, color transform, and gamma encoding. Results for our method and ground truth image are shown: (a) our CNN-based ISP and (b) ground truth image. In each image, a zoomed in view of the white rectangular region is displayed in the top left hand corner (inside the black rectangle). It can be seen that the CNN output looks almost identical to the ground truth image.

## ABSTRACT

A conventional camera performs various signal processing steps sequentially to reconstruct an image from a raw Bayer image. When performing these processing in multiple stages the residual error from each stage accumulates in the image and degrades the quality of the final reconstructed image. In this paper, we present a fully convolutional neural network (CNN) to perform defect pixel correction, denoising, white balancing, exposure correction, demosaicing, color transform, and gamma encoding. To our knowledge, this is the first CNN trained end-to-end to perform the entire image signal processing pipeline in a camera. The neural network was trained using a large image database of raw Bayer images. Through extensive experiments, we show that the proposed CNN based image signal processing system performs better than the conventional signal processing pipelines that perform the processing sequentially.

## 1 Introduction

An image signal processing (ISP) pipeline is important when reconstructing an image from raw Bayer image for display applications. In a conventional camera, dedicated hardware is employed to perform image signal processing in a modular architecture. There are various processing steps performed in a conventional ISP pipeline to reconstruct an image faithfully. The main processes performed in an ISP include denoising, white balancing, exposure correction, demosaicing, color transform, and gamma encoding.

Generally, color filters are placed on top of the silicon photodetectors to capture a scene at different wavelength ranges to reproduce its color. Bayer color filter array (CFA) is widely used in consumer cameras. Bayer mosaic contains four pixel elements with red, blue and two green filter elements placed in a 2X2 pixel grid. Demosaicing is performed to interpolate the missing red, green, or blue values in the Bayer color filter array. When recording a scene there are various sources of noise that corrupt the recorded signal. Example noise sources include dark signal nonuniformity, photon shot noise, and read out noise. Some of these noise sources are additive while others are multiplicative. The denoising step is implemented in an ISP to reduce the noise in the signal. As a photodetector has a limited charge well capacity, a scene with high dynamic range luminance variation will make the charge well to overflow or underflow. For example, the brighter regions will make the charge well to overflow while the darker regions such as shadow regions will make the charge well to underflow. This may lead to visible artifacts in the reconstructed image. To account for the extreme luminance variation in a scene, the charge integration time (exposure time) is adjusted according to the luminance level of the scene. The exposure correction is performed to account for the variation in charge integration time of an image sensor when capturing a scene. The human visual system exhibits a phenomenon known as 'color constancy' to discount the illuminant effect on the perceived color of a scene. To mimic the function of human color constancy, a white balancing step is implemented in a camera image processing pipeline. White balancing removes the illuminant color from the image sensor response and transforms the image to look as if it was captured under a white light such as D65 (daylight illuminant with correlated color temperature 6500K). Since the response function of a camera does not perfectly match the color matching functions of the human visual system, the image sensor responses are transformed to a standard color space that represents the recorded color independent of the characteristics of the imaging device. This is an important step to communicate color between devices and to correctly reproduce color for display applications. The color conversion step is implemented in an ISP to transform the device dependent color responses to a device independent color representation model such as sRGB. The human visual system responds nonlinearly to linear variation of scene luminance. However, most cameras have approximately linear response to luminance variation. Gamma encoding is performed to account for the mismatch between the luminance response function of the human visual system and that of a camera. Further, gamma encoding also helps to compress more data using a limited number of bits by compressing high luminance regions in the same way as the human visual system.

Many of the processes performed in an ISP pipeline are ill-posed problems, so it is impossible to find a closed form solution. To overcome this problem, conventional modular based algorithms apply hand-crafted heurists-based approaches to derive a solution independent of the rest of the processing in an ISP pipeline. Many of the modular based methods independently make assumptions about the scene or sensor or both to derive a hand-crafted solution. However, these assumptions do not hold in uncontrolled outdoor and indoor environments. Therefore, the reconstructed image quality degrades with real world images.

Sequentially performing various ISP processes using modular based algorithms poses another major challenge as the residual error from each processing module accumulates in the reconstructed signal. In particular, the later stages have to correct for the intended processing and the residual error left in the signal by the previous modules in the ISP pipeline. This degrades the quality of the reconstructed image. However, performing multiple processing in one-step or using a convolutional neural network (CNN) to perform all the stages in an ISP reduces artifacts (example: color moiré and zippering) and accumulation of error in the reconstructed signal compared to the conventional modular based ISPs. The main reason for error accumulation in the conventional ISP is that each module uses a task-specific loss function independent of the other modules. Due to the mismatch in the loss functions used in different processing modules, the accumulated error increases as we progress through a conventional ISP pipeline. However, a CNN based approach uses a single loss function to optimize the entire processing involved in an ISP pipeline in an end-to-end optimization setting. Therefore, the optimization minimizes the loss function that measures the reconstruction error in the final output image to achieve a better quality image.

## 2 Related work

In the past, many different modular based approaches have been proposed to perform various processing steps involved in an ISP [Buchsbaum 1980; Malvar et al. 2004; Alleysson 2005; Buades et al. 2005; Lu et al. 2010]. These methods perform one of the processing in an ISP pipeline based on some assumptions about the scene or the image sensor. For example, Buchsbaum [1980] proposed an algorithm for illuminant estimation based on the assumption that the arithmetic mean of a scene color is achromatic. However, this assumption does not always hold in real world scenes. For example, the algorithm fails when there is dominant color present in a scene or a single colored object occupies a large region of a scene. Land and McCann [1971] proposed a well-known algorithm called the 'Retinex' for white balancing. This algorithm considers the highest value in each color channel (RGB) as the white representation in an image to estimate the illuminant color of the scene. However, using a single or a few pixels in a scene may not give reliable estimate for the illuminant color due to noise. Further, specular regions could cause the image sensor to saturate and lead to incorrect estimate for the illuminant. Cheng et al. [2014] proposed an algorithm for illuminant correction in an image by applying principal component analysis on the color distribution of a scene. Finlayson and Trezzi [2004] proposed an algorithm for illumination estimation based on the color statistics of the scene. In this algorithm, the authors used Minkowski norm to estimate the illuminant. Based on the grey-edge hypothesis, Weijer et al. [2007] proposed an algorithm for illuminant estimation. In this algorithm, the authors assumed that the average color difference between pixels in a scene is achromatic. Recently, convolutional neural network based solutions have been proposed for illumination correction and shown to be successful compared to conventional methods [Barron 2015; Bianco et al. 2015; Lou 2015; Qian et al. 2017].

Demosaicing has been widely researched in the past and various methods have been proposed including edge-preserving interpolation schemes [Li et al. 2008], nonlinear filter-banks [Dubois 2006], channel correlations based approach [Chang et al. 2015], median filtering [Hirakawa and Parks 2005], luminance channel interpolation [Zhang et al. 2009], and methods that utilize self-similarity and redundancy properties in natural images [Buades et al. 2009; Menon and Calvagno 2011]. A number of different approaches has been proposed using conventional methods and neural network based methods [Zhang et al. 2009b; Buades et al. 2005; Liu et al. 2013; Patil and Rajwade 2016; Akiyama et al. 2015 ]. There are recent works that propose convolutional neural network based solutions for denoising [Jain and Seung 2009; Shahdoosti and Rahemi 2019; Zhang et al. 2017; Lefkimmiatis 2018; Zhang et al. 2017; Romano et al. 2017; Burger et al. 2012], demosaicing [Syu et al. 2018; Kokkinos and Lefkimmiatis 2018; Syu et al. 2018], debluring [Schuler et al. 2016], and image enhancement [Dong et al. 2016; Kim et al. 2016; Ahn at al. 2018; Bigdeli et al. 2017]. These authors showed that the convolutional neural network based methods to provide better results than the conventional methods.

Although, there are modular based solutions for various processing involved in an ISP pipeline there is no clear order identified to perform these modular processing. Kalevo and Rantanen [16], investigated in which order demosaicing and denoising should be performed in an ISP pipeline. Based on their empirical evidence they concluded that denoising is to be performed before demosaicing. Zhang et al. [2009] argued that performing demosaicing before denoising will generate noise-caused color artifacts in the demosaiced image. However, there are effective methods that perform demosaicing before denoising [Zhang et al. 2009]. To overcome this ordering confusion of which process to perform first, recent methods propose to perform demosaicing and denoising both together in a single step or in a single algorithm and are shown to perform better than performing in separate modules [Gharbi et al. 2016; Klatzer et al. 2016; Schwartz et al. 2018]. Recently, a CNN has been proposed for joint denoising and demosaicing by Gharbi et al. [2016]. Their network takes a Bayer image and noise level in the image as inputs to jointly perform denoising and demosaicing. To train the network, the authors mined millions of Internet images to collect the hard image regions and used these image regions to train their network. Although the network performs denoising and demosaicing together, it requires calculating the noise level in the input image in advance and adding it to the input image as an additional layer. With real world image sensors, it is not possible to model the noise accurately. Schwartz et al. [2018] proposed a CNN to perform demosaicing, denoising and image enhancement together. Though the authors claimed that the neural network learned how to perform this processing, the input to the network was already demosaiced using bilinear interpolation. Therefore, the network operates not on the raw sensor data but on already demosaiced data.

A space-varying filter based approach has been proposed for joint denoising and demosaicing by Menon and Calvagno [2009]. The authors formulate the demosaicing problem as a linear system and performed denoising on the color and luminance components separately. Zhang et al [2009] proposed a joint denoising and demosaicing algorithm based on spatially adaptive principal component analysis on the raw image sensor data. Their method exploits the spatial and spectral correlations in a CFA image to remove the noise while maintaining the high frequency color edges in the image. However, the spatial and spectral correlations do not hold for both natural and artificial scenes [Farinella et al. 2008]. Heide et al. [2014] developed a framework to perform common image processing steps in an ISP based on the natural-image priors. We would like to note that the natural-image priors do not hold for all the scenes, and therefore, leads to degradation in image quality. The authors formulated the image reconstruction as a linear least-squares problem with non-linear regularizers. They applied nonlinear optimization algorithms to find an optimal solution using proximal operators. Recently, a generative adversarial network has been proposed to perform joint demosaicing and denoising using perceptual optimization [Dong et al. 2018]. Zhou et al. [2018] proposed a residual neural network for joint demosaicing and super resolution by performing an end-to-end mapping between Bayer images and high-resolution RGB images. They showed that performing multiple processing in a single step reduces errors and artifacts that are common when performed separately. Zhao al. [2017] investigated various loss functions for image restoration. Other methods perform joint demosaicing and denoising include methods proposed by Hirakawa and Parks [2006], Khashabi et al. [2014], Klatzer et al. [2016], Paliy et al. [2008], Condat [2010], Menon and Calvagno [2009], Goossens et al. [2013], Paliy et al. [2008], Hirakawa [2008], Zhou et al. [2018], Fang et al. [2012], Klatzer et al. [2016], Condat and Mosaddegh [2012], and Henz et al. [2018].

The above described classical and CNN based solutions perform either individual process or a combination of two processes at most in an ISP pipeline. However, there is no deep CNN based method proposed to replace the entire ISP pipeline yet. Motivated by the prior works that perform more than one ISP processing in a single module, we propose a fully convolutional deep neural network to perform several image signal processing steps, including defect pixel correction, denoising, white balancing, exposure correction, demosaicing, color transform, and gamma encoding by feeding raw Bayer image sensor data as an input to the network and training the network end-to-end using a single loss function. We demonstrate qualitatively and quantitatively that our neural network based ISP performs better than the existing methods.

**Contributions:**

- We developed a novel CNN model to perform image signal processing in a camera.

- We presented the first CNN to perform the entire ISP pipeline including defect pixel correction, denoising, white balancing, exposure correction (low light and high light level correction), demosaicing, color transform, and gamma encoding

- We will release the raw Bayer image data in the public domain.

- Performed more realistic test for image denoising using our method and state of the art methods. In particular, we tested these methods with both additive and multiplicative noise.

- We showed that the proposed CNN-based ISP pipeline can work with other CFA mosaics such as X-Trans by Fujifilm.

## 3 CNN for image signal processing

Traditionally ISP pipelines have been implemented as sequential processing steps using a bank of linear or nonlinear filters based on some assumptions about the statistical distribution of color in an image. This sequential processing has been shown to accumulate error as the image progresses through the pipeline and leads to poor image quality [Zhou et al. 2018]. Recently, CNN has been shown to be successful in performing various computer vision and image processing tasks [Krizhevskyet al. 2012; Simonyan and Zisserman 2014; Szegedy et al. 2015; He et al. 2016; Gharbi et al. 2016]. The advantage of using a CNN to implement the entire ISP pipeline is that the parameters of the CNN can be optimized in an end-to-end manner by minimizing a single loss function that carefully measures the accuracy of the reconstructed output image.

## 3.1 Network Architecture

Figure 2 illustrates the neural network architecture that we used to implement ISP pipeline. Our neural network configurations are quite different from the conventional neural networks. In particular, we pass the short connections through a convolutional layer. This helped our network to learn the entire processing (defect pixel correction, denoising, white balancing, exposure correction, demosaicing, color transform, and gamma encoding) involved in an ISP pipeline with relatively a small network. In the Microsoft ResNet [He et al. 2016] architecture, the residual learning block performs identical mapping of the input to the output of the block. This simple residual block did not give us satisfactory performance; since the residual blocks make an identical copy of the input to the output, the network did not learn to generalize the complex ISP pipeline. However, the authors of ResNet were able to achieve better performance for object detection/recognition by naively stacking many residual blocks to the network. Compared to ResNet we are using a significantly less number of layers. Further, performing the entire ISP processing using a fewer number of convolutional layers is challenging and we cannot afford to have residual blocks that perform identical copy of the input. Empirically, we found that feeding the parallel connections (short connections) through a convolutional layer improved the performance of the neural network. The network consists of four parallel connections with one main path and three short connections. To match the dimensions of the layer to which the short connection is concatenated, two of the short connections were first processed with 2X2 average pooling (stride=2). However, the main path was processed with 2X2 max pooling (stride=2). This was performed to get the advantage of both pooling methods when reconstructing an image. Max pooling has been widely used for object recognition applications [Simonyan and Zisserman 2014]. However, max pooling may not be the best for reconstruction applications. Therefore, we used average pooling for the short connections to capture the first order statistics of the activation from each activation region. Based on Schwartz et al. [2019] we used tanh nonlinearity in all three short connections after performing batch normalization. Each parallel connection is concatenated to the main path followed by a 1X1 convolution to reduce the depth of the concatenated layer to 64. Except 1X1 convolutional layers, all the other convolutional layers were performed with 3X3 kernels with stride of 1. The convolutional layers were created by convolving with 64 filter kernels (however, output layer used only 3 kernels to produce RGB image). Input to convolution layers were padded to maintain the output to have the same dimensions as the input.

Motivated by the VGGnet [Simonyan and Zisserman 2014] and U-net [Ronneberger et al. 2015] architectures, we perform 2X2 max pooling with stride of 2 to reduce the input size in the main path. However, we do not increase the depth of the layers as the spatial dimension is reduced. This was performed to force the network to find a compact latent representation of the raw sensor data while preserving the important information about the scene to correctly reconstruct the image at the output layer. We performed up sampling to bring the dimensions of the hidden representation back to the input dimensions. All the 3X3 convolutional layers in the main path were followed by a batch normalization and a LeakyReLu nonlinear activation function except the output layer. The output layer has no batch normalization but, uses a sigmoid function to ensure the reconstructed image is bounded between 0 and 1.

## 3.2 Loss function

To obtain the best performance, it is not enough to have the best network architecture but also important to have the appropriate loss function that accurately measures the perceptual quality of an image. Reconstruction of a raw sensor image into an RGB image can be formulated as follows:

$$y = f(x) + n \quad (1)$$

where $x \in \mathbb{R}^N$ denotes the reconstructed RGB image, $y \in \mathbb{R}^N$ denotes the observed raw Bayer CFA image data, and $n$ denotes the noise from various sources. The function $f(.)$ is the degradation function that models the quantum efficiency of the silicon, response of the readout circuit, and the CFA transfer function. To make this problem simple, $f(.)$ is generally assumed to be a linear function and replaced with an N dimensional square matrix or a diagonal matrix [Kokkinos and Lefkimmiatis 2018]. Other than the responses of photo detector, CFA pattern, and read-out circuit; the measured response is also corrupted by noise from various sources including dark response of the photo detectors, fixed pattern noise from the readout circuit and photo detector irregularities in the sensor array, and photon shot noise. Shot noise is generally modelled as Poisson distribution. Given that there are many unknowns, finding a closed form solution to $x$ is an ill-posed problem. In the past, a number of algorithms have been proposed by assuming simple linear models or assumptions about the statistical color distribution of an image. Here, we treat the problem as a nonlinear inverse estimation problem and use a carefully designed CNN to find an optimal estimate for x. A well-known method to formulate this problem is to apply Bayes rule and maximize the posterior probability as follows:

$$P(x/y) \propto P(y/x) * P(x) \quad (2)$$

where $P(y/x)$ is the likelihood term, $P(x)$ is the prior probability on x. To obtain the best estimate for $x$, we need to maximize the posterior probability $P(x/y)$. Taking logarithm to both sides of equation (2) results,

$$argmax_x\{\log(P(x/y)\} = argmax_x\{log(P(y/x)) + log(P(x)))\} \quad (3)$$

More formally, the MAP estimation in equation (3) can be expressed as an optimization problem as follows:

$$\hat{x} = argmin_x \left\{ \begin{array}{c} \alpha * \|y - f(x)\|^2 \\ +(1-\alpha) * DOG(x) * \|y - f(x)\| \end{array} \right\} \quad (4)$$

here, $\hat{x}$ is the reconstructed image. The negative log-likelihood term can be written as $\|y - f(x)\|^2$ and the negative log-prior term (regularizer term) can be written as $DOG(x) * \|y - f(x)\|$. In this expression $DOG(x)$ denotes difference of Gaussian of the reconstructed image. Through experimentation, we found that modeling the regularizer term as a weighted L1 norm worked better for preserving the high frequency edges in an image. We weighted the likelihood term and the regularizer term using a weighting term α. We used α = 0.9 in our optimization.

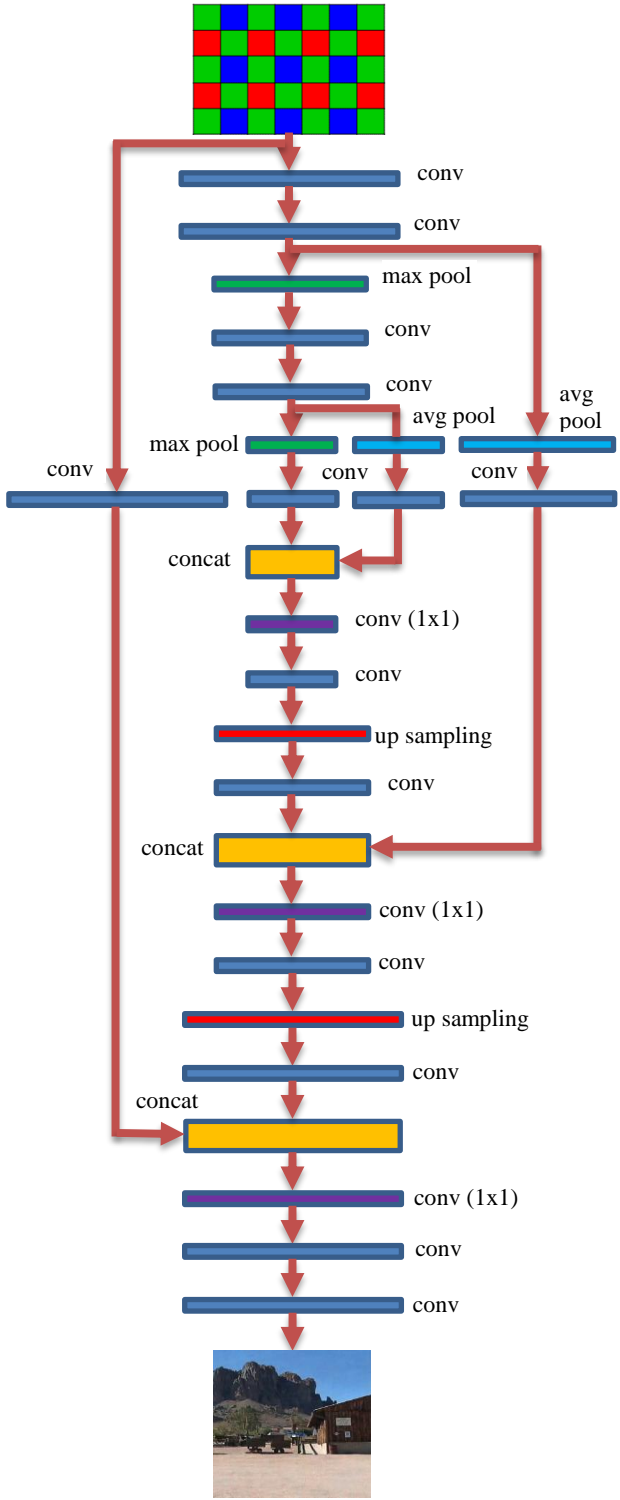

**Figure 2: Proposed neural network to perform image signal processing to reconstruct the RGB image from Bayer sensor data. The network consists of a main path and three skip connections. All the skip connections were processed with a convolutional layer for better image reconstruction.**

### 3.3 Data set

It has been shown that the Kodak data set [Li et al. 2008] and McMaster data set [Zhang et al. 2011] do not represent the real world image statistics [Dong et al. 2018; Syu et al. 2018]. In addition, these two data sets have only 24 and 18 images respectively. In this paper, we used a much larger data set of 11347 images with ground truth illuminants [Ciurea and Funt 2003]. As our CNN performs the entire ISP pipeline including white balance correction, Ciurea and Funt [2003] data set is more appropriate as we can test our CNN for illumination correction. However, this is not possible with the Kodak or McMaster data sets.

### 3.4 Noise modeling

Recorded image sensor response is corrupted by various sources of noise. Due to random variation of detected photons in an image sensor, the image sensor response is corrupted by photon shot noise. In modern cameras, the pixel size is reduced to increase the resolution of the camera. However, photon noise increases as the pixel size is reduced [Blanksby et al. 1997]. Currently, photon noise is the most significant type of noise in an image sensor system that degrades the image quality [Blanksby et al. 1997]. This noise component is signal dependent and very different from additive white Gaussian noise widely used in the literature when evaluating demosaicing and denoising algorithms [Foi et al. 2008]. In our evaluations, we modelled the photon noise as a signal dependent noise component and modelled it separately from other sources of noise for realistic evaluation of our CNN and competing methods.

The read-out noise arises due to electronic inefficiencies in reading the accumulated charge and converting the electrical charge into a digital pixel value. Image sensor response is affected by both additive noise and multiplicative noise [Lukáš et al. 2006]. For example, Photo Response Non Uniformity (PRNU) noise is a multiplicative noise whereas fixed pattern noise is additive noise [Lukáš et al. 2006]. However, in the past, many of the demosaicing and denoising algorithms were evaluated with additive noise only [Gharbi et al. 2016, Heide et al. 2014, and Kokkinos and Lefkimmiatis 2018]. For a more realistic investigation of our CNN-based ISP and the competing methods, we modelled both additive noise and multiplicative noise in an image capturing system and incorporated them into our reverse imaging pipeline.

### 3.5 Generation of Bayer image data

The raw Bayer CFA images were generated from a database of images [Ciurea and Funt 2003]. This image set contains RGB images and the ground truth illuminant. We used our in-house inverse ISP pipeline built based on one of our CMOS image sensor models to create the Bayer data from the RGB images. First, the inverse pipeline linearizes the RGB image by removing the gamma encoding and represents the linearized image with a higher precision than the input RGB image. Then we convert the sRGB to device dependent space using a transformation matrix obtained from one of our sensors. The device dependent RGB responses were then rendered using our inverse pipeline to simulate three different exposure conditions (long, medium, and short) and the out of range pixels were clipped. Shot noise was modelled as multiplicative noise with two different SNR levels 25dB and 30dB. Fixed pattern noise from various sources was modelled as additive Gaussian noise. However to simulate the irregularities along the rows and columns in an image sensor response, we used 2D sinusoidal waves in row and column directions with zero mean Gaussian noise overlaid on the 2D sinusoidal patterns. This approximately models the fixed-pattern noise variation due to irregularities in the silicon photoreceptors, and read-out noise along the column and row pixel elements. Finally, the image was run through a Bayer mosaic simulator to generate a Bayer CFA image. With two different noise levels and three different integration times, we were able to generate 6 images from each of the original RGB images. In each image, a gray ball was placed (the ball was fitted on the camera) to obtain the ground truth illuminant. We cropped the images to remove the gray ball to avoid the neural network learning to perform white balance correction and exposure correction based on the gray ball. In particular, we took four different crops of 240X220 pixels image. This created 272000 raw images of different noise levels and different exposure conditions (low light and high light images). We split the images by randomly assigning the images to training (240000), test (16000) and validation (16000) sets.

To generate ground truth images for each of the corresponding raw Bayer images, we took the linearized images and performed illumination correction using the ground truth illuminant obtained from the gray ball measurements. This image was gamma encoded to obtain the ground truth image to train our CNN.

### 3.6 Training

We trained our neural network end-to-end using the raw CFA image responses as input and the corresponding ground truth images as the target output. The network was implemented in Keras with Tensorflow backend [Chollet 2015]. We used the Adam optimizer with a starting learning rate 0.001 with other parameters kept as default. The Adam optimizer is a flavor of a stochastic gradient descent algorithm that also takes advantage of the Root Mean Square Propagation (RMSProp) and Adaptive Gradient (AdaGrad) algorithms [Kingma et al. 2014]. We used a batch size of 32 and minimum learning rate to 0.000001. During the training process, we halve the learning rate if the loss calculated on the validation set did not improve for 100 epochs. This was required to reach the optimum point in the space spanned by the loss function. In our training and testing, we kept the image size to 240X220 pixels. The filter weights were initialized using random uniform distribution. Training was performed on a NVIDIA quadro P5000 GPU and Intel® Xeon® w2175 CPU.

## 4 Performance evaluation

In this section, we compare the performance of our CNN-based ISP to other existing modular based approaches. As there is no single algorithm proposed to perform the entire ISP pipeline, we compare our CNN with existing methods that perform single processing or multiple processing, such as joint demosaicing and denoising. For a fair comparison, we used the ground truth estimates to perform the missing processes of the competing methods. For example, if a competing method performs only denoising and demosaicing, we performed the rest of the processing, such as white balance correction, and gamma encoding, using the ground truth values.

### 4.1 Results for white balancing

We compared the performance of our neural network based ISP for color constancy with well-known color constancy algorithms. The results are listed in Table 1. Angular error has been widely used to measure the performance of color constancy algorithms [Hordley and Finlayson 2004]. Therefore, we calculated the mean angular error between the ground truth illuminant and the illuminant estimated by each of the algorithms in the RGB space. We perform quantitative and qualitative comparison of our neural network with the following algorithms: white patch [Land and McCann 1971], gray world [Buchsbaum 1980], gray edge [Weijer et al. 2007], weighted gray edge [Gijsenij et al. 2012], PCA based algorithm [Cheng et al. 2014], and shades of gray [Finlayson and Trezzi 2004]. As we have discussed in the previous section, each of these algorithms makes assumptions about the color variation in a scene to estimate the illuminant. However, these assumptions do not hold for all the natural and artificial scenes. From the results reported in Table 1, we can see that our CNN-based ISP performs better than the rest of the methods and the PCA-based algorithm provides the least performance. It can also be seen that gray world, shades of gray and gray edge algorithms provide a comparable performance. This is because these three algorithms estimate the illumination based on the computation of Minkoviski norm given by:

$$\left(\frac{\int (f(x))^p dx}{\int dx}\right)^{\frac{1}{p}} = ke \quad (5)$$

where p is the order of the norm. For $p = 1$ the equation becomes gray world assumption, for $p = 6$ the equation becomes the shades of gray and with L1 norm the equation becomes gray edge hypothesis [Finlayson and Trezzi 2004].

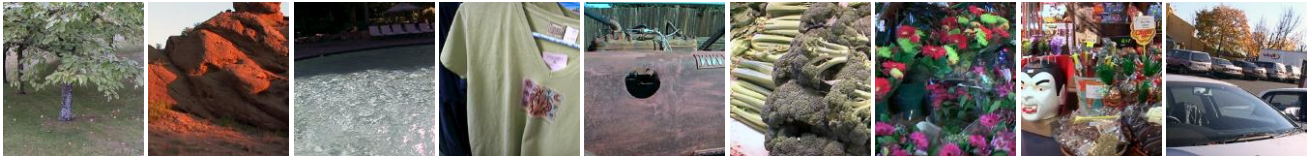
(b) White patch [Land and McCann 1971]

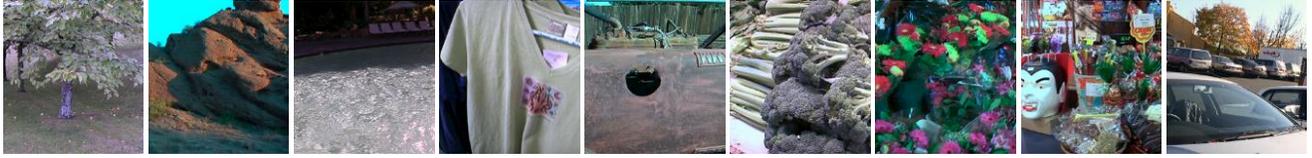
(c) Gray world [Buchsbaum 1980]

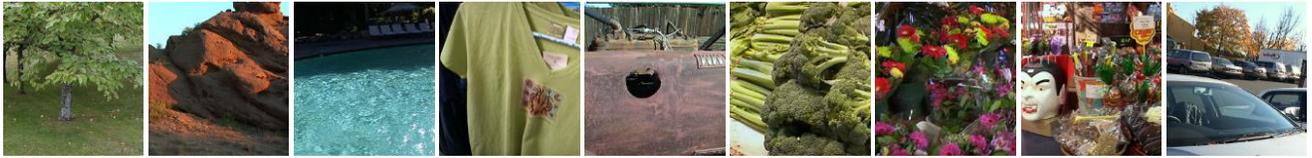
(d) Gray edge [Weijer et al. 2007]

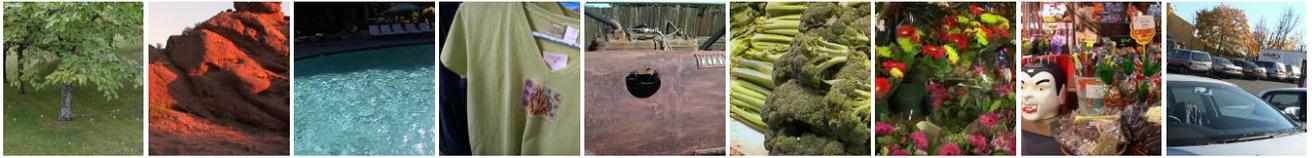
(e) Weighted gray edge [Gijsenij et al. 2012]

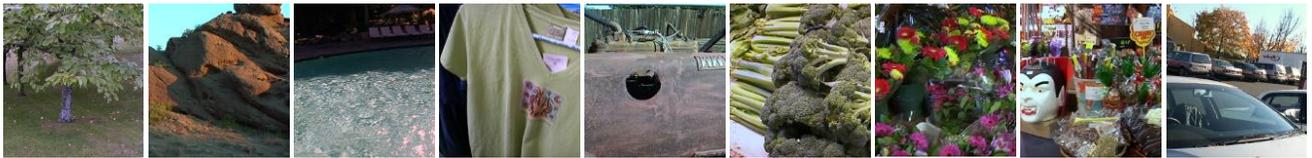
(f) Shades of gray [Finlayson and Trezzi 2004]

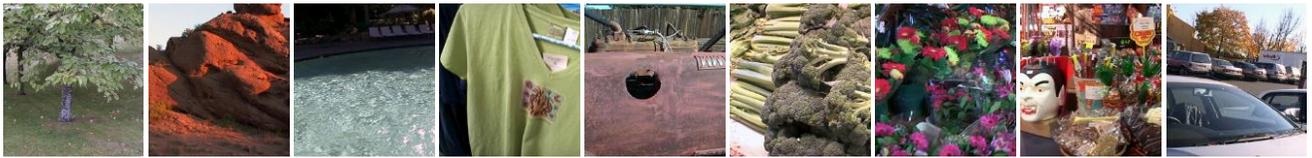
(j) PCA based [Cheng et al. 2014]

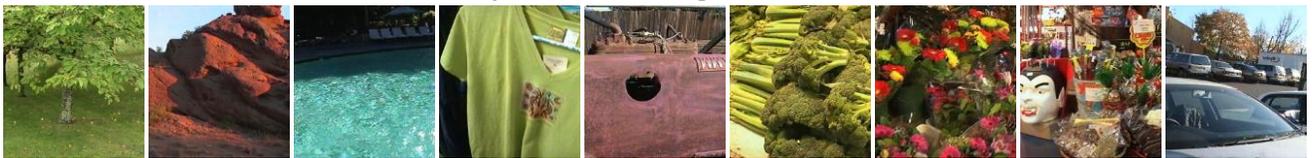
(h) Our CNN

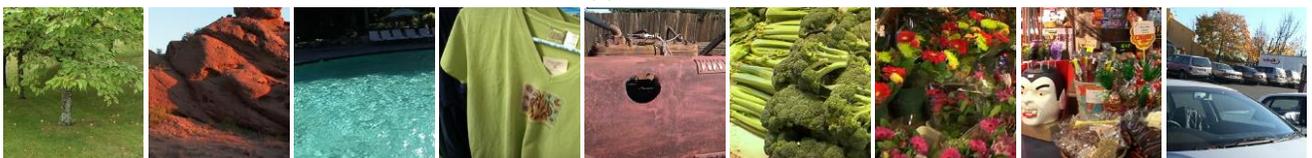
(i) Ground truth

**Figure 3:** Color constancy results for our CNN based ISP and competing methods. Our CNN performs consistently better than the other methods.

| Algorithm | Angular error |
|---|---|
| White patch [Land and McCann 1971] | 6.7 |
| Gray world [Buchsbaum 1980] | 3.6 |
| Gray edge [Weijer et al. 2007] | 4.3 |
| Weighted gray edge [Gijsenij et al. 2012] | 6.5 |
| PCA based [Cheng et al. 2014] | 10.9 |
| Shades of gray [Finlayson and Trezzi 2004] | 4.4 |
| Our CNN | 2.8 |

**Table 1:** *Experimental results for our CNN and other methods for color constancy. Left hand column lists the methods used in the evaluation and the right hand column lists the angular error (degrees) calculated on the 16000 test images.*

Example results for color constancy is shown in Figure 3. It can be seen that our CNN-based ISP performs the best. Notably, if the scene has a dominant color, many of the algorithms give poor performance (see columns 1 to 6 in Figure 3). However, if the image has many different colors, the algorithms perform relatively better (see columns 7 to 8 in Figure 3). Further, the performance of the algorithms improve if the scene contains white objects (see columns 9 in Figure 3). Our CNN-based approach performs consistently better regardless of the color content of the scene.

### 4.2 Results for image reconstruction

In this section we compare the performance of our CNN-based ISP pipeline with other demosaicing, denoising, and image enhancing algorithms. In particular, we compare the performance of our CNN with the following algorithms: bilinear interpolation, FlexISP by Heide et al. [2014], Tan et al. [2017], Malvar et al. [2004], Lu et al. [2010], Zhang et al. [2009], Menon et al. [2007], Su [2006], and Jeon and Dubois [2013]. Example images from each of these methods are shown in Figure 1 and Figure 4. Peak signal to noise ratio (PSNR) and mean signal to noise ratio (SNR) for each of these algorithms tested on the 16000 test images are listed in Table 2. From these results, it can be seen that our CNN-based approach performs better than other methods. The bilinear interpolation together with Wiener filter for noise removal performs the least.

### 4.3 Results for defective pixel correction

In imaging devices, defective pixels are pixels that do not sense light levels correctly. A defective pixel could be a dead pixel or a pixel that has light sensitivity that is significantly high or low compared to the rest of the pixel array (stuck pixels). Defective pixels in an image sensor can occur due to various reasons including short circuit, dark current leakage, and damage or debris in the optical path. To simulate defect pixels in an image sensor array, we randomly made 0.01% of the pixel responses to either 0 or 255. We trained our CNN to learn to identify and correct the response of the defective pixels. Results for defect pixel correction are shown Figure 5. From these results it can be seen that our CNN-based ISP pipeline can effectively perform defect pixel correction.

### 4.4 Results for other color filter mosaics

To investigate how our CNN-based ISP pipeline performs with other color filter mosaics, we trained our CNN using X-Trans color filter mosaic by Fujifilm. The X-Trans color filter mosaic has 6X6 pattern of photosites. In a 6X6 cell array, X-Trans has more green filter elements compared to the standard Bayer filter mosaic. Test results are shown in Figure 6. From these results, it can be seen that our CNN-based ISP pipeline can be easily adapted to other nonstandard color filter mosaics as well.

| Algorithm | PSNR | Mean SNR |
|---|---|---|
| FlexISP [Heide et al. 2014] | 21.31 | 14.45 |
| ADMM [Tan et al. 2017] | 20.92 | 13.91 |
| Malvar et al. [2004] | 21.52 | 14.66 |
| Lu et al. [2010] | 28.64 | 21.78 |
| Zhang et al. [2009] | 25.57 | 18.72 |
| Menon et al. [2007] | 29.72 | 22.88 |
| Su [2006] | 29.76 | 22.91 |
| Jeon and Dubois [2013] | 26.91 | 20.06 |
| Bilinear interpolation, Wiener filter | 18.02 | 11.17 |
| Our CNN | 30.71 | 24.58 |

**Table 2:** *Performance comparison of our CNN-based ISP and other existing methods. First column lists the method, second column lists the PSNR and the third column lists the mean SNR calculated on the 16000 test images.*

### 4.5 Network configuration

We have experimented with different network configurations including plain encoder-decoder pair, identical copy in the short connections like ResNet [He et al. 2016] or U-net [Ronneberger et al. 2015] and found that passing the short connections through a convolutional layer provided better reconstruction results than the other configurations we tested. All the results reported in this paper are based on the network shown in Figure 2. We also experimented with different depth for the hidden layers and found that reducing the depth from 64 to 32 or smaller value increases the PSNR of the reconstructed image. The network shown in Figure 2 requires 438k weight parameters and takes 215ms to reconstruct an image (240X220 pixels) on our system with Intel® Xeon® w2175 CPU.

### 4.6 Limitations and future work

As we used supervised learning to train our CNN-based ISP, it relies on the training data to learn the processing involved in an ISP pipeline. However, if the data is not representative of a given problem or if the ground truth data is corrupted with noise and/or artifacts, the network will learn to produce the noise and artifacts that are in the training data. Therefore, the success of a data driven method depends on the training data. An alternative way to train a network is to use an unsupervised method or partially supervised method such as reinforcement learning or a generator-discriminator pair (example: generative adversarial network). Another possible future direction is that expanding the functions of the network to other processing such as motion blur, super resolution, and high dynamic range imaging. A more interesting direction would be to develop a neural network that learns to restore an image corrupted by an unknown degradation function.

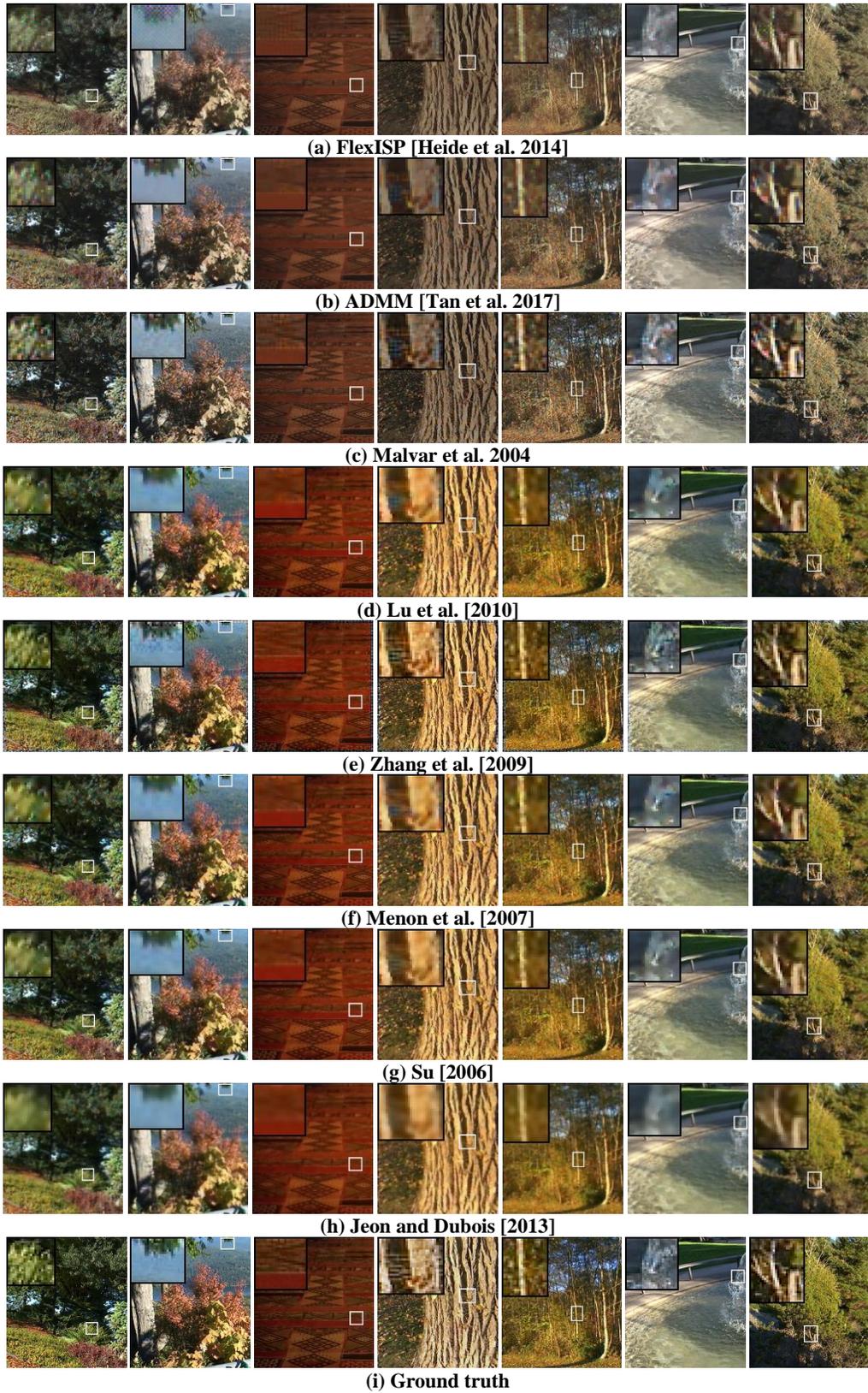

Figure 4: Image reconstruction results for existing methods and ground truth image (results for our CNN are shown in Figure 1).

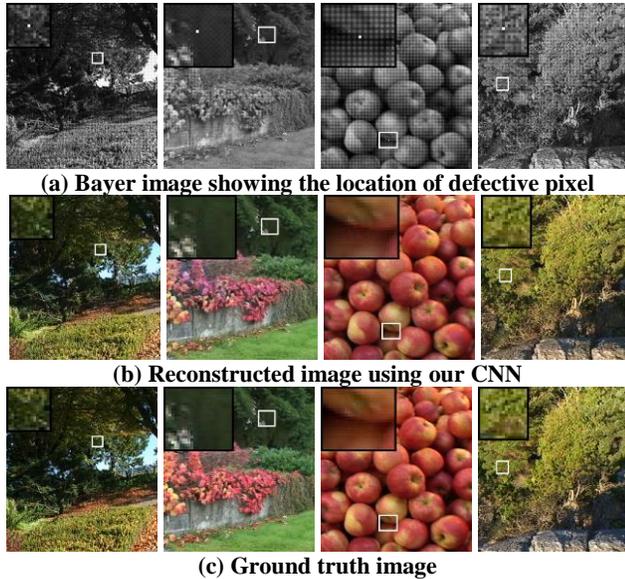

**(a) Bayer image showing the location of defective pixel**

**(b) Reconstructed image using our CNN**

**(c) Ground truth image**

**Figure 5: Test results for defect pixel correction: (a) shows the location of the defective pixel with a white dot inside the white rectangle. A zoomed in view of the defect pixel region is shown in the top left hand corner in (a), (b) and (c).**

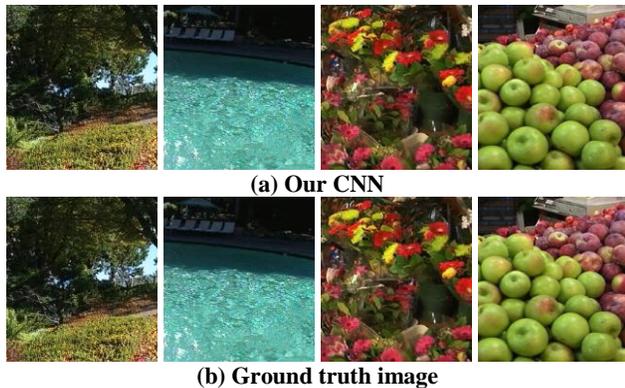

**(a) Our CNN**

**(b) Ground truth image**

**Figure 6: Reconstruction results for X-Trans CFA by Fujifilm.**

## 5 Conclusions

We developed a CNN based image signal processing pipeline for performing defect pixel correction, denoising, white balancing, exposure correction, demosaicing, color transform, and gamma encoding. We demonstrated that performing the entire image processing steps using a CNN performs better than the conventional modular based approaches including methods that jointly perform demosaicing and denoising. We have illustrated quantitative and qualitative results for our CNN and other existing methods and shown that our CNN-based ISP performs better under challenging conditions.